\documentclass[manuscript]{aastex61}

\newcommand\iras{IRAS 15398$-$3359 }
\newcommand\sm{$M_\odot$}
\received{}
\revised{}
\accepted{\today}
\submitjournal{}

\begin{document}
\title{The co-evolution of disk and star in embedded stages: the case of the very low-mass protostar \iras
}

\email{}

\author[0000-0002-0786-7307]{YUKI OKODA}

\author{YOKO OYA}
\affiliation{Department of Physics, The University of Tokyo, 7-3-1, Hongo, Bunkyo-ku, Tokyo 113-0033, Japan; okoda@taurus.phys.s.u-tokyo.ac.jp}

\author{NAMI SAKAI}
\affiliation{RIKEN Cluster for Pioneering Research, Wako, Saitama 351-0198, Japan}

\author{YOSHIMASA WATANABE}
\affiliation{Department of Physics, The University of Tsukuba, Tsukuba, Ibaraki 305-8577, Japan}
\affiliation{Tomonaga Center for the History of the Universe, Faculty of Pure and Applied Sciences, University of Tsukuba,  Tsukuba, Ibaraki 305-8571, Japan}

\author{JES K. J\O RGENSEN}
\affiliation{Centre for Star and Planet Formation, Niels Bohr Institute and Natural History of Denmark, University of Copenhagen, \O ster Voldgade 5-7, DK-1350, Copenhagen K, Denmark}

\author{EWINE F. VAN DISHOECK}
\affiliation{Leiden Observatory, Leiden University, PO Box 9513, 2300 RA, Leiden, The Netherlands Max-Plank Institut f$\ddot{u}$r Extrsterrestrische Physik (MPE), Giessenbachstr.1, 85748, Garching, Germany}

\author{SATOSHI YAMAMOTO}
\affiliation{Department of Physics, The University of Tokyo, 7-3-1, Hongo, Bunkyo-ku, Tokyo 113-0033, Japan; okoda@taurus.phys.s.u-tokyo.ac.jp}


\begin{abstract}
\par We have observed the CCH ($N=3-2, J=7/2-5/2, F=4-3\ \rm and\ 3-2$) and SO ($6_7-5_6$) emission at a 0\farcs2 angular resolution toward the low-mass Class 0 protostellar source \iras with ALMA. The CCH emission traces the infalling-rotating envelope near the protostar with the outflow cavity extended along the northeast-southwest axis. On the  other hand, the SO emission has a compact distribution around the protostar. The CCH emission is relatively weak at the continuum peak position, while the SO emission has a sharp peak there.  
Although the maximum velocity shift of the CCH emission is about 1 km s$^{-1}$ from the systemic velocity, a velocity shift higher than 2 km s$^{-1}$ is seen for the SO emission. 
 This high velocity component is most likely associated with the Keplerian rotation around the protostar. The protostellar mass is estimated to be 0.007$^{+0.004}_{-0.003}$ \sm\ from the velocity profile of the SO emission. 
With this protostellar mass, the velocity structure of the CCH emission can be explained by the model of the infalling-rotating envelope, where the radius of the centrifugal barrier is estimated to be 40 au from the comparison with the model.
The disk mass evaluated from the dust continuum emission by assuming the dust temperature of 20 K$\sbond$100 K is 0.1$\sbond$0.9 times the stellar mass, resulting in the Toomre Q parameter of 0.4$\sbond$5.
Hence, the disk structure may be partly unstable.
All these results suggest that a rotationally-supported disk can be formed in the earliest stages of the protostellar evolution.

\end{abstract}




\section{Introduction} \label{sec:intro}
\par It is generally thought that a rotationally-supported disk is formed around a newborn star, which can eventually evolve into a planetary system \citep*[e.g.,][]{Bodenheimer(1995), Saigo&Tomisaka(2006), Machida et al.(2016)}. A thorough understanding of disk formation is therefore of fundamental importance in exploring the origin of the Solar System. Such disks are indeed found in many protostellar objects in the Class II and Class III stages \citep*[e.g.,][]{Williams and Cieza (2011), Dutrey et al.(2014)}.
Although their existence was suggested for a few protostars in the younger stages (Class 0/I)  \citep*[e.g.,][]{Enoch et al.(2009)}, it has recently been established, thanks to high spatial resolution observations with interferometers including the Atacama Large Millimeter/Submillimeter Array (ALMA) \citep*[e.g.,][]{Tobin et al.(2012), Yen et al.(2013), Yen et al.(2017), Brinch&Jorgensen(2013), Murillo et al.(2013), Lindberg et al.(2014), Lee et al.(2014), Ohashi et al.(2014), Oya et al.(2016), Oya et al.(2017), Sakai et al.(2017)}. These results suggest that the disk structure is possibly formed at an earlier stage than previously thought. Although theoretical simulations have extensively been conducted for disk formation \citep*[e.g.,][]{Bate1998, Hueso&Gullot(2005), Inutsuka et al.(2010), Machida et al.(2011), Tsukamoto et al.(2017), Zhao et al.(2018)}, our understanding is far from complete. Thus, disk formation is one of the frontiers in star-formation studies.
Further through Kepler's law it is possible to estimate protostellar masses, even when the young star itself cannot be observed \citep*[e.g.,][]{Lommen et al.(2008)}.
Thus, studies of disk kinematics make it possible to trace the amount of material that has been accreted onto the young star and thus follow its build up during the embedded protostellar stages.

\par \iras is a low-mass protostar in the Lupus 1 molecular cloud at a distance of 155 pc \citep{Lombardi et al.(2008)}. Its bolometric temperature is 44 K \citep{Jorgensen et al.(2013)}, typically of Class 0 protostars.
A molecular outflow from this source was detected through single-dish observations of its CO emission \citep{Tachihara et al.(1996), van Kempen et al.(2009)}. From a chemical point of view, \iras is a so-called warm carbon-chain chemistry (WCCC) source which is rich in various unsaturated carbon-chain molecules, such as CCH, $\rm C_{4}$H, and C$\rm H_3$CCH, present on scales of a few thousand au scale around the protostar \citep{Sakai et al.(2009)}.

\par \cite {Jorgensen et al.(2013)} conducted sub-arcsecond resolution observations toward \iras with ALMA, and found a ring structure of the H$^{13}\rm CO^{+}$ ($J=4-3$) emission at scales of 150$-$200 au. This ring structure is thought to be formed by the destruction of HCO$^{+}$ through the gas-phase reaction with H$_2$O which has sublimated from the grains due to the enhanced protostellar luminosity caused by a recent accretion burst. \cite {Bjerkeli et al.(2016b)} observed the HDO ($1_{0,1}-0_{0,0}$) emission with ALMA, and found that it is localized at the cavity wall in the vicinity of the protostar. The extent of the emission is also consistent with the interpretation that a recent accretion burst has taken place.
In addition, \cite {Bjerkeli et al.(2016a)} find bullets with spacings consistent with the time-scale for relatively recent accretion bursts.

\par \cite {Oya et al.(2014)} reported the distribution of the $\rm H_2$CO ($5_{1,5}-4_{1,4}$) line toward this source with ALMA at high angular resolution of 0\farcs5 ($\sim$80 au), and characterized a bipolar outflow extending along the northeast-southwest axis at about a 2000 au scale. 
The analysis of the outflow structure indicates that it is oriented almost in the plane of the sky with an inclination angle of 70\degr\ with respect to the line of sight.
This in turn suggests that the disk/envelope system is seen edge-on.
This feature is further verified by CO observations with the Submillimeter Array (SMA) \citep{Bjerkeli et al.(2016a)}. \cite{Oya et al.(2014)} derived an upper limit on the protostellar mass to be 0.09 $M_\odot$ from the velocity structure of the H$_2$CO emission.
In the envelope, they pointed out the possibility that a rotationally-supported disk structure may have already been formed around the protostar based on the detection of  high velocity components of $\rm H_2$CO associated with the protostar whose velocity shift is as high as $\sim$3 km s$^{-1}$.
Recently, \cite{Yen et al.(2017)} also found an upper limit of 0.01 $M_\odot$ for the protostellar mass, based on their observation of the $\rm C^{18}$O ($J=2-1$) line (resolution of about 0\farcs5).

\par From these results, it seems likely that \iras has a very low protostellar mass. However, all values obtained so far are upper limits, and the mass has not been evaluated definitively.
Thus, we need to confirm the existence of the disk component around the protostar suggested by \cite{Oya et al.(2014)} and use it to measure the protostellar mass.
With these motivations, we conducted the ALMA observations at a higher angular resolution of 0\farcs2  ($\sim$ 30 au) to characterize the disk/envelope structure of IRAS 15398$-$3359.

\section{Observation} \label{sec:style}
\par Observations of \iras were carried out with ALMA in its Cycle 2 operation on 2015 July 20. 
Spectral lines of CCH and SO were observed with the Band 6 receiver at frequency range from 261 to 263 GHz. The spectral line parameters are listed in Table \ref{Table}. 
The $N=4-3$ lines of CCH used by \cite{Oya et al.(2014)} do not clearly show the velocity structures because of mutual overlapping of the two hyperfine components, and hence, we here chose the $N=3-2$ lines having a larger separation of the hyperfine components.
Forty-one antennas were used in the observations, where the baseline length ranged from 14.90 to 1559.16 m.
The field center of the observations was ($\alpha_{2000}$, $\delta_{2000}$)= (15\fh43\fm02\fs242, $-$34\arcdeg 09\arcmin 06\farcs70). 
The system temperature ranged from 60 to 100 K during the observation.
The backend correlator was set to a resolution of 61 kHz and a bandwidth of 59 MHz.
This spectral resolution corresponds to a velocity resolution of 0.07 km s$^{-1}$ at 250 GHz.
The bandpass calibrator was J1517$-$2422. 
The data calibration was performed in the antenna-based manner, and the absolute calibration accuracy is 10 \% \citep{Lundgen.2013}, where the absolute flux density scale was derived from Titan.  Images were prepared by using the CLEAN algorithm, where Briggs' weighting with a robustness parameter of 0.5 was employed. 
The continuum image was obtained by averaging line-free channels, and the line images were obtained after subtracting the continuum component directly from the visibilities. 
Self-calibration was not applied in this study, since the continuum emission is not bright enough.
The primary beam (half-power beam) width was 23\farcs04.
The total on-source time was 21.61 minutes.
The synthesized-beam sizes are 0\farcs21$\times$0\farcs15 (P.A. 58\arcdeg) for the continuum image, 0\farcs36$\times$0\farcs29 (P.A. 60\arcdeg) for the CCH image, and 0\farcs22$\times$0\farcs16 (P.A. 55\arcdeg) for the SO image.
The rms noise levels for the continuum, the CCH emission, and the SO emission are 0.12 mJy bea$\rm m^{-1}$, 4 mJy bea$\rm m^{-1}$, and 4 mJy bea$\rm m^{-1}$, respectively, for the channel width of 61 kHz.\\
 
\section{Results and Discussions} \label{sec:floats}
\subsection {Distribution}
\par Figure \ref{Figure1}(a) shows the moment 0 map of the CCH ($N=3-2, J=7/2-5/2, F=4-3\ \rm and\ 3-2$) lines, where the extended outflow cavity along the northeast-southwest axis (P.A. 220\arcdeg) is clearly seen. This feature is essentially similar to those for the CCH ($N=4-3, J=7/2-5/2, F=4-3\ \rm and\ 3-2$) lines reported by \cite{Jorgensen et al.(2013)} and \cite{Oya et al.(2014)}. On 300 au scales in the vicinity of the protostar, the CCH emission is extended along the northwest-southeast axis.
Figures \ref{Figure1}(b) and (c) are blow-ups of the central part of panel (a).
In Figure \ref{Figure1}(b), the continuum map is superposed with contours on the CCH map. The coordinates of the peak position and its intensity are derived from a 2D Gaussian fit to the image : ($\alpha_{2000}$, $\delta_{2000}$) = (15\fh43\fm02\fs2421$\pm0.0002$, \sbond34\arcdeg 09\arcmin 06\farcs805$\pm0.001$). The position is consistent with previous reports \citep{Jorgensen et al.(2013), Oya et al.(2014), Bjerkeli et al.(2016b)}. The peak intensity of 6.98$\pm$0.12 mJy bea$\rm m^{-1}$ at 1.2 mm is also consistent with previous observations  at 0.8 mm (19 mJy bea$\rm m^{-1}$ with a beam size of about 0\farcs55$\times$0\farcs37; P. A. $-82\arcdeg$) \citep{Jorgensen et al.(2013)} and 0.6mm (36 mJy bea$\rm m^{-1}$ with a beam size of about 0\farcs55$\times$0\farcs37; P. A. $-82\arcdeg$) \citep{Bjerkeli et al.(2016b)}assuming optically thin emission from dust with the opacity law $\kappa \propto \nu^{\beta}$ with $\beta \simeq $ 2 typical for standard interstellar medium dust.
\par The CCH emission is weak at the continuum peak position, as shown in Figure \ref{Figure1}(b).
We define the line perpendicular to the outflow axis as the envelope direction (P.A. 130\degr; the arrow in Figure \ref{Figure1}(b)), and prepare the intensity profiles of the CCH, SO, and 1.2 mm continuum emission along the line (Figure \ref{Figure1}(d)). 
Apparently, CCH shows a double peak in its intensity profile.
The intensity peak of CCH appears on both sides at a distance of about 70 au from the continuum peak.
Although this double-peaked feature was marginally reported in the CCH ($N=4-3, J=7/2-5/2, F=4-3\ \rm and\ 3-2$) emission by \cite{Jorgensen et al.(2013)} and \cite{Oya et al.(2014)}, it is further confirmed in the present high-resolution observation.
\par In contrast to the CCH distribution, the SO distribution is concentrated in the vicinity of the protostar.
Figure \ref{Figure1}(c) shows the integrated intensity map of the SO ($6_7-5_6$) line in contours superposed on that of CCH in a color map.
The peak position of the SO distribution almost coincides with the continuum peak. This feature is clearly seen in the intensity profile along the envelope direction, as shown in Figure \ref{Figure1}(d), where the SO emission shows a single-peaked distribution between the two intensity peaks of CCH.

\par A similar difference between the CCH and SO distributions is reported for another Class 0 low-mass protostar, L1527. The gas distribution around the protostar of L1527, which is a prototypical WCCC source, was explored with ALMA by \cite{Sakai et al.(2014a), Sakai et al.(2014b)}. According to their results, CCH and c-$\rm C_3$$\rm H_2$ are distributed outside the centrifugal barrier of the infalling-rotating envelope, while SO resides at the centrifugal barrier and/or inside it.
A centrifugal barrier stands for the perihelion of the ballistic motion of the infalling-rotating gas, where the gas cannot fall inward under the conservation laws of the energy and angular momentum.
It has been proposed that the temperature is raised around the centrifugal barrier due to a weak accretion shock of the infalling gas, and SO is likely liberated from grain mantles and enhanced there \citep{Sakai et al.(2014a), Sakai et al.(2014b), Aota et al.(2015), Miura et al.(2017)}.
In contrast, CCH seems to be broken up by gas phase reactions or depleted onto grain mantles inside the centrifugal barrier.
\subsection{Kinematics} 
\subsubsection{Keplerian Motion} 
\par First, we investigate the kinematic structure of the SO emission, which is well concentrated around the protostar. Figure \ref{Figure2}(a) shows the position-velocity (PV) diagram of the SO emission along the envelope direction (P.A. 130\degr) centered at the continuum peak. 
Red-shifted and blue-shifted components can be recognized in the southeastern and northwestern parts, respectively (Figure \ref{Figure2}(a)).
More importantly, the maximum velocity shift from the systemic velocity is as high as about 3 km s$^{-1}$.
These components likely correspond to the high velocity components marginally detected in the $\rm H_2$CO emission \citep{Oya et al.(2014)}.
These results imply that SO traces the rotating disk structure around the protostar. 
The PV diagram along the line perpendicular to the envelope direction does not show a significant velocity gradient, indicating no infall  and outflow motion (Figure \ref{Figure2}(b)). Hence, the observed rotational motion is most likely the Keplerian rotation. 
In Figure \ref{Figure2}(a), the blue contours represent the model of Keplerian rotation with a protostellar mass of 0.007 $M_\odot$.
It seems to explain the velocity structure of the PV diagram observed for the SO emission reasonably well. Note that the intensity around the central position in the model is much stronger than that in the observed PV diagram. This is most likely due to self-absorption or absorption by the foreground gas, especially around the systemic velocity. In this study, we focus on the kinematic structure of the disk, and the effects of radiative transfer and self-absorption are not included in the model.
On the other hand, the PV diagram cannot be explained by the infalling-rotating envelope, even if the radius of the centrifugal barrier is set to 1 au (almost the free-fall motion) (Figure \ref{Figure2}(c)).
The counter velocity components, which are red-shifted and blue-shifted for the northwestern and southeastern sides, respectively, appear in this model as opposed to the observed PV diagram.
\par Thus, the protostellar mass of this source is evaluated to be only 0.007$^{+0.004}_{-0.003}$ \sm, assuming Kepler rotation (Figure \ref{Figure2}(d)).
This very small value is consistent with the upper limits reported so far: $<$ 0.09 $M_\odot$ \citep{Oya et al.(2014)}; $<$ 0.01 $M_\odot$ \citep{Yen et al.(2017)}. 
Although \iras clearly has a very low protostellar mass, a rotating disk structure has already formed around the protostar.
Note that, in the above analysis, we employ a systemic velocity of 5.5 km s$^{-1}$.
This is slightly different from the value (5.0$-$5.3 km s$^{-1}$) reported in \cite {Oya et al.(2014)} and \cite{Yen et al.(2017)}. This difference is discussed later.
\par The protostellar mass derived above is comparable to, or even smaller than, the mass of the first hydrostatic core expected from star formation theories \citep*[0.01-0.05 \sm:][]{Penston(1969), Larson(1969), Masunaga et al.(1998), Saigo&Tomisaka(2006)}. Since the protostellar mass is so small, the gravity of the system may not be well represented by a central force field. In fact, the dust mass evaluated from the 1.2 mm continuum data is between 0.006 and 0.001 $M_\odot$ for an assumed dust temperature of 20 K and 100 K, respectively, which are not much different from the protostellar mass. Here, we {\bf employ} the mass absorption coefficient of 6.8 $\times$ 10$^{-4}$ cm$^{2}$ g$^{-1}$ at 1.2 mm \citep{Ward-thompson et al.(2000)}.
Although the temperature of 100 K seems too high, it is adopted as the upper limit.
Hence, the rotation motion may not be exactly Keplerian, and the mass of 0.007 $M_\odot$ might be an apparent value. Nevertheless, the small protostellar mass would not change drastically, even if such an effect is considered \citep{Mestel.1963}. 
To explore this effect in more detail and to derive the protostellar mass definitively, we need higher sensitivity observations of SO and other molecules.
Even if the motion can be approximated by the Keplerian motion, a caveat should be mentioned.
With the current resolution, we cannot be sure whether the materials in the beam centered at the protostar has already accreted onto the star.
In this case, the derived mass could be an upper limit to the mass of the prototstar.
\par To assess the stability of the disk, we evaluated the Toomre-Q parameter \citep{Toomre 1964, Goldreich&Lynden-Bell(1965)} by assuming Keplerian motion.
The derived Q parameter is 0.4 and 5 for temperatures of 20 K and 100 K, respectively.
Thus, either the disk is relatively warm and of low mass, or it is in the unstable regime.
Such an unstable part may be responsible for future accretion bursts.


\subsubsection{Infalling-rotating Envelope}
\par By using the protostellar mass estimated from the Keplerian motion, we examine whether the kinematic structure traced by the CCH emission is consistent with an infalling-rotating envelope.
Figure \ref{Figure3}(a) is the PV diagram of the CCH emission along the envelope direction, while Figure \ref{Figure3}(b) shows that along the line perpendicular to the envelope direction. The dashed lines are the model of the Keplerian rotation with the protostellar mass of 0.007 $M_\odot$. 
The Keplerian rotation model does not explain the velocity gradient along the line perpendicular to the envelope direction (Figure \ref{Figure3}(b); dashed lines). The contours in Figures \ref{Figure3}(a) and (b) are the results of the infalling-rotating envelope model \citep{Oya et al.(2014)} around the protostar with a protostellar mass of 0.007 $M_\odot$, where the radius of the centrifugal barrier is assumed to be at 40 au.
The infalling-rotating motion seems to roughly explain the observed kinematic structure of CCH, although there is weak emission outside of the model in the observed PV diagrams probably due to contributions from the outflow cavity.
It should be noted that, in the infalling-rotating envelope case, the systemic velocity of 5.3 km s$^{-1}$, which is similar to the previous report \citep{Yen et al.(2017)}, gives a better fit.
Hence, the envelope and the disk could have slightly different systemic velocities.
This difference is small, but significant.
It may originate from the small protostellar mass, because, in this case, the disk system and the envelope could have different centers of mass.

\subsection{Disk around a Very Low-Mass Protostar}
\par \iras is found to be a very low-mass protostar. A protostar with a mass as low as this source has never been reported. Nevertheless, we find that a rotating disk structure has already formed (Figure \ref{Figure2}(a)). 
Although the radius of centrifugal barrier is derived to be 40 au, SO could become abundant in front of the centrifugal barrier due to the accretion shock. Hence, its distribution up to a radius of about 60 au is reasonable.
The Keplerian rotation fit is mostly done in the range of $r \la$ 40 au, and hence, the effect of the infalling-rotating components to the derivation of the protostellar mass can be ignored.
\cite{Kristensen et al.(2012)} and \cite{Jorgensen et al.(2013)} reported envelope masses of 0.5 $M_\odot$ and 1.2 \sm, respectively, and hence, the protostar will grow further.
Thus, the very low mass of the protostar means that it is in its infancy.
In fact, the dynamical timescale of the outflow of this source is reported to be 10$^{2}-10^{3}$ yr \citep {Oya et al.(2014), Bjerkeli et al.(2016a)}.
The present results therefore means that a rotating disk structure can be formed at a very infant stage of protostellar evolution.
\par Figures \ref{Figure4} (a) and (b) show the comparison of the protostellar masses with the bolometric luminosties and the disk masses, respectively. 
Red marks represent the result for \iras obtained in this study. 
Compared with some other low-mass protostars reported in previous works, the mass of \iras is the lowest, while its bolometric luminosity is moderate. This result indicates that the bolometric luminosity is not always a good indicator of the protostellar mass.
On the other hand, the disk mass is the lowest among the sources in Figure 4(b). The disk mass seems to increase with increasing protostellar mass, although more systematic studies are necessary.


\par The mass accretion rate $\dot{M}_{\rm acc}$ averaged over the protostar life is estimated to be about $\sim7\times10^{-6}$ $M_\odot \rm\ yr^{-1}$ from the protostellar mass of 0.007 $M_\odot$ and the above dynamical timescale. 
It can also be estimated by using of the following relation \citep{Palla(1991)}:
\[
\dot{M}_{\rm acc}=\frac{LR_{\rm star}}{GM},
\]
where $L$ is the luminosity and $R_{\rm star}$ the radius of the protostar.\ We evaluate the current accretion rate $\dot{M}_{\rm acc}\ \rm to\  be \ 2.1\times 10^{-5}$ $M_\odot$ yr$^{-1}$ by using $L$ of 1.8 $L_\odot$ \citep{Jorgensen et al.(2013)} and $R_{\rm star}$ of 2.5 $R_\odot$ \citep*[e.g.,][]{Palla1999, Baraffe2010}. 
It should be noted that we roughly employ the bolometric luminosity for $L$ and the average radius of the protostar for $R_{\rm star}$.
The average and current accretion rates are almost within a range of cannonical values $10^{-5}-10^{-6}\ M_\odot \rm\ yr^{-1}$ \citep{Hartmann et al.(1997)}.
It should be noted that the mass loss rate due to the outflow is reported to be (3.2 $-\ 3.7) \times 10^{-6}$ $M_\odot$ yr $^{-1}$ \citep{Yildiz et al.(2015)} and 7 $\times\ 10^{-8}$ $M_\odot$ yr $^{-1}$ \citep{Bjerkeli et al.(2016a)}, which is comparable to or smaller than the above estimates of the accretion rate.
\par Although the very low mass of this protostar can naturally be interpreted as its infancy, the alternative possibility is that this protostar may evolve into a brown dwarf.
A planetary system could be formed even around a brown dwarf, as predicted by theoretical models for the formation of Earth-like planets around a brown dwarf \citep{Payne and Lodato(2007)}.
However, this interpretation may be not the case for this source, because the envelope mass around the protostar is as high as 0.5$-$1.2$ M_\odot$ \citep{Kristensen et al.(2012), Jorgensen et al.(2013)}.
In any case, the observational characterization of very low-mass protostars is important for further understandings of the stellar and disk formation process and its diversity. More sensitive observations are awaited in this direction.
\\
\par 
This paper makes use of the following ALMA data set:
ADS/JAO.ALMA\#2013.1.01157.S. ALMA is a partnership of the ESO (representing its member states), the NSF (USA) and NINS (Japan), together with the NRC (Canada) and the NSC and ASIAA (Taiwan), in cooperation with the Republic of Chile.
The Joint ALMA Observatory is operated by the ESO, the AUI/
NRAO, and the NAOJ. 
The authors are grateful to the ALMA staff for their excellent support.
This study is supported by Grant-in-Aid
from the Ministry of Education, Culture, Sports, Science, and
Technologies of Japan (25108005 and 18H05222).
JKJ acknowledges support from the European Research Council (ERC) under the
European Union's Horizon 2020 research and innovation programme (grant agreement No. 646908) through ERC Consolidator Grant ''S4F''. Research at Centre for Star and Planet Formation is funded by the Danish National Research Foundation.
EvD is supported by EU A-ERC grant
291141 CHEMPLAN.
Yuki Okoda thanks the Advanced
Leading Graduate Course for Photon Science (ALPS)
for financial support.




\begin{figure}[h]
\centering
\includegraphics[scale=0.5]{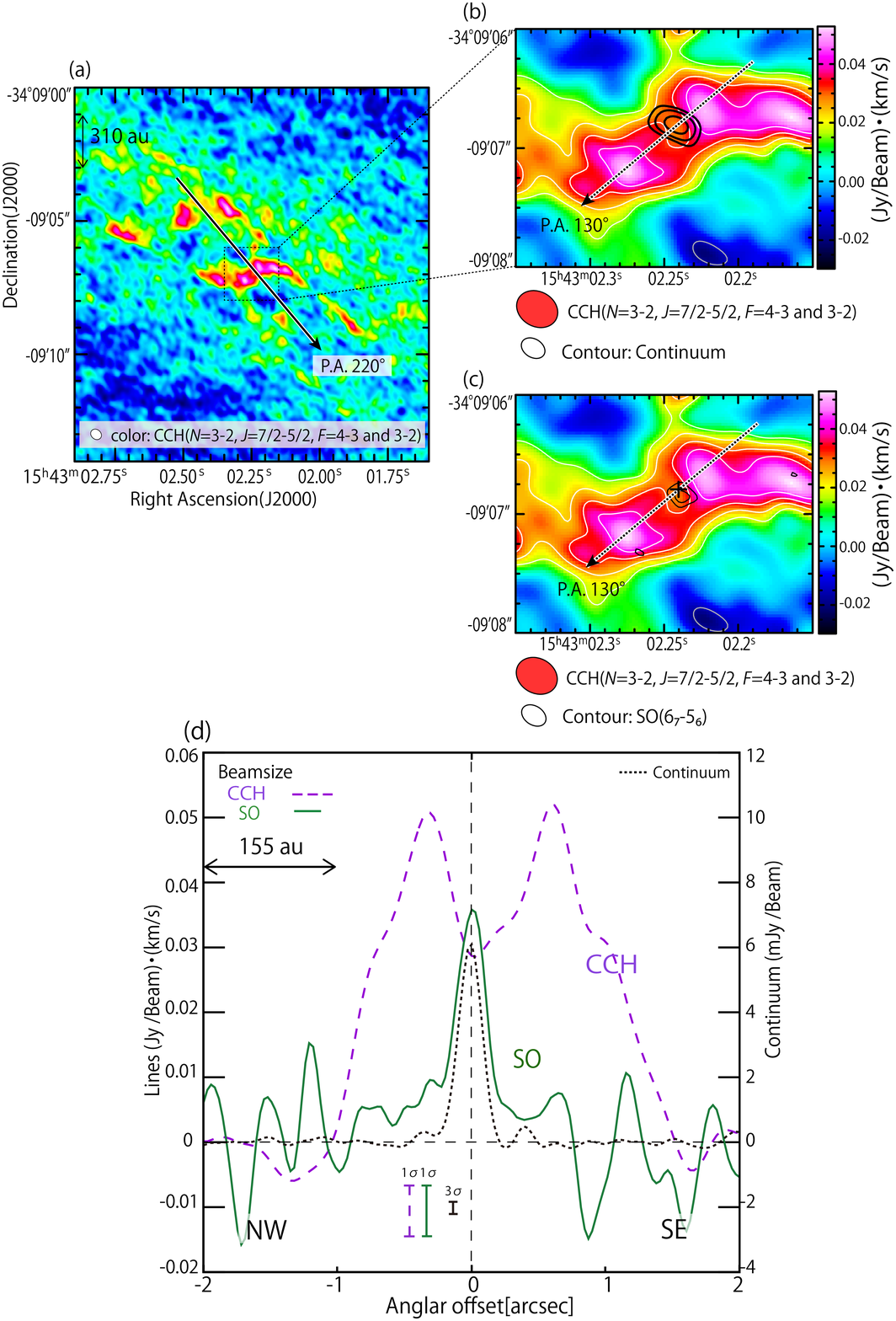}
\end{figure}

\begin{figure}[h]
\caption{(a) The moment 0 map of the CCH ($N=3-2, J=7/2-5/2, F=4-3\ \rm and\ 3-2$) lines. The arrow represents the outflow axis (P.A. 220\degr). (b) The color map is a blow-up of the central part of panel (a). The black contours are the continuum map contour levels are 10$\sigma$, 20$\sigma$, and 40$\sigma$, where $\sigma$ is 0.12 mJy beam$^{-1}$. The white contours are the moment 0 map of the CCH ($N=3-2, J=7/2-5/2, F=4-3\ \rm and\ 3-2$) lines. Contours levels are 2$\sigma$, 3$\sigma$, 4$\sigma$, 5$\sigma$, and 6$\sigma$, where $\sigma$ is 8 mJy beam$^{-1}$ km s$^{-1}$. The dashed arrow represents the envelope direction (P.A. 130\degr). (c) The color map is a blow-up of the central part of panel (a). The contours are the moment 0 map of the SO ($6_7-5_6$) line. Contour levels are 3$\sigma$, 4$\sigma$, where $\sigma$ is 8 mJy beam$^{-1}$ km s$^{-1}$. The black cross shows the continuum peak position.  (d) The intensity profiles of CCH ($N=3-2, J=7/2-5/2, F=4-3\ \rm and\ 3-2$) and SO ($6_7-5_6$) along the envelope direction shown by the dashed arrow in panel (b). 
Their noise levels are described by $\sigma$, where $\sigma$ is 8 mJy beam$^{-1}$ km s$^{-1}$, on the bottom of this figure. 
The noise level of the continuum is described by 3$\sigma$, where $\sigma$ is 0.12 mJy beam$^{-1}$.
The abscissa is the angular offset from the continuum peak. 
\label{Figure1}}

\end{figure}

\begin{table}[h]
\centering
\caption{Parameters of the Observed Line \label{Table}}
\begin{tabular}{cccccc}
\hline \hline
 & Transition & Frequency\ (GHz) & $S \mu^2$($D^2$) & $E_{\rm u}$$k^{-1}(\rm K)$ & Beam size\\
 \hline
 CCH & $N=3-2, J=7/2-5/2, F=4-3$ & 262.0042600 & 2.3
 & 25 & 0\farcs36 $\times$ 0\farcs29 (P.A. 60\degr) \\
 & $N=3-2, J=7/2-5/2, F=3-2$ & 262.0064820 & 1.7
 & 25 &\\ 
 \hline
 SO & $6_{7}-5_{6}$ & 261.8437210 & 16.4
 & 47 & 0\farcs22 $\times$ 0\farcs16 (P.A. 55\degr)\\
\hline
\end{tabular}
\end{table}

\begin{figure}[h]
\centering
\includegraphics[scale=0.8]{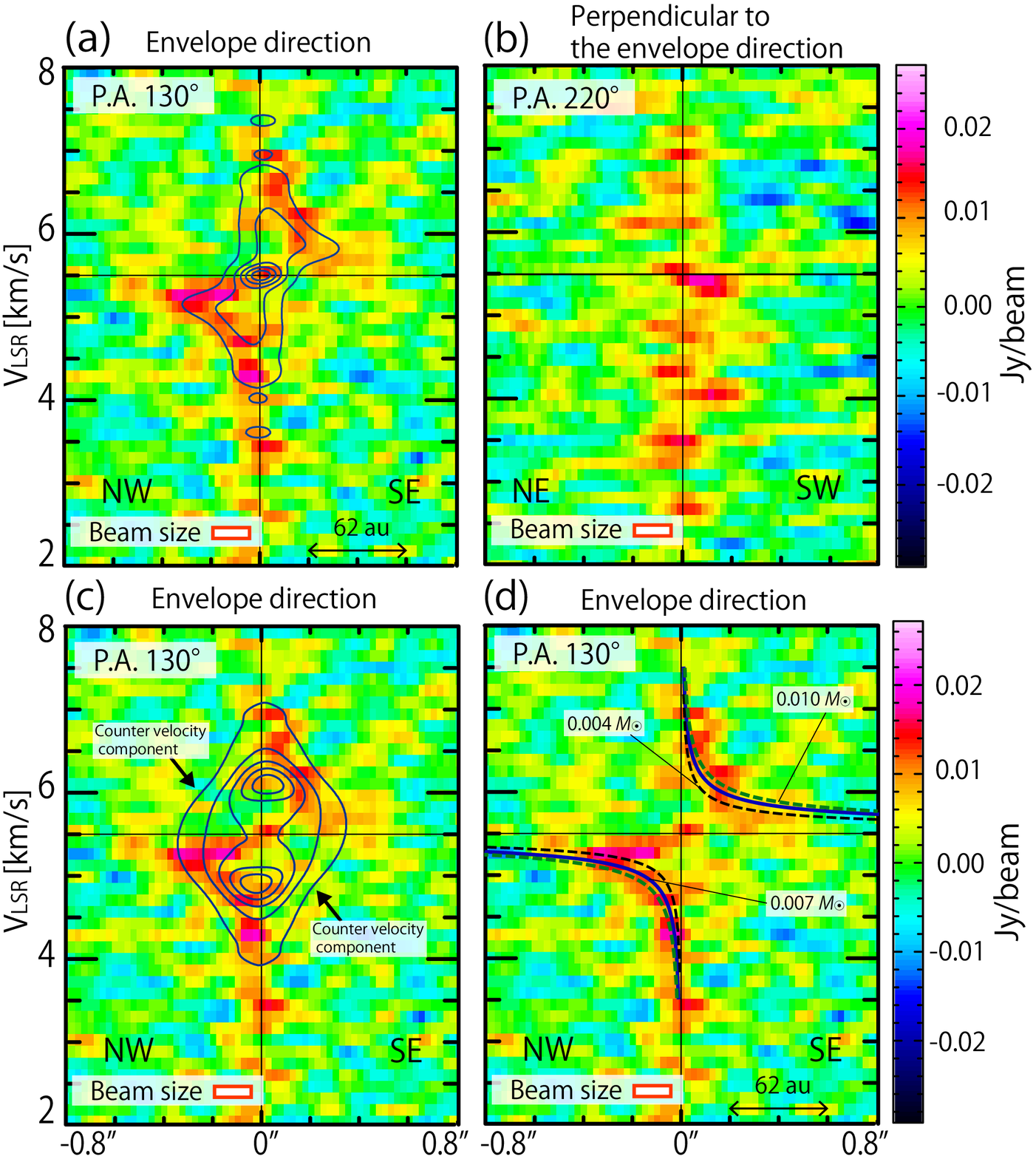}
\end{figure}

\begin{figure}[h]

\caption{(a) The PV diagram of the SO ($6_7-5_6$) emission along the envelope direction shown by the dashed arrow in Figure 1(b). The origin is the continuum peak position. The blue contours are the model of Keplerian rotation around the systemic velocity of 5.5 km s$^{-1}$ with the protostellar mass of 0.007 $M_\odot$, the outer radius of 40 au, and the inclination angle of 70\degr (0\degr for a face-on configuration). (b) The PV diagram of the SO ($6_7-5_6$) emission along the line perpendicular to the envelope direction centered at the continuum peak position. (c,d) The color maps are the PV diagrams of the SO ($6_7-5_6$) emission along the envelope direction. The blue contours are the infalling-rotating envelope model with the protostellar mass of 0.007 $M_\odot$, the outer radius of 40 au, the radius of the centrifugal barrier of 1 au, and the inclination angle of 70\degr.
The curves are the model of Keplerian rotation with the different protostellar mass assumed. Green, blue and black show the cases of 0.010, 0.007 and 0.004 $M_\odot$. \label{Figure2}}

\end{figure}

\begin{figure}[h]
\centering
\includegraphics[scale=0.7]{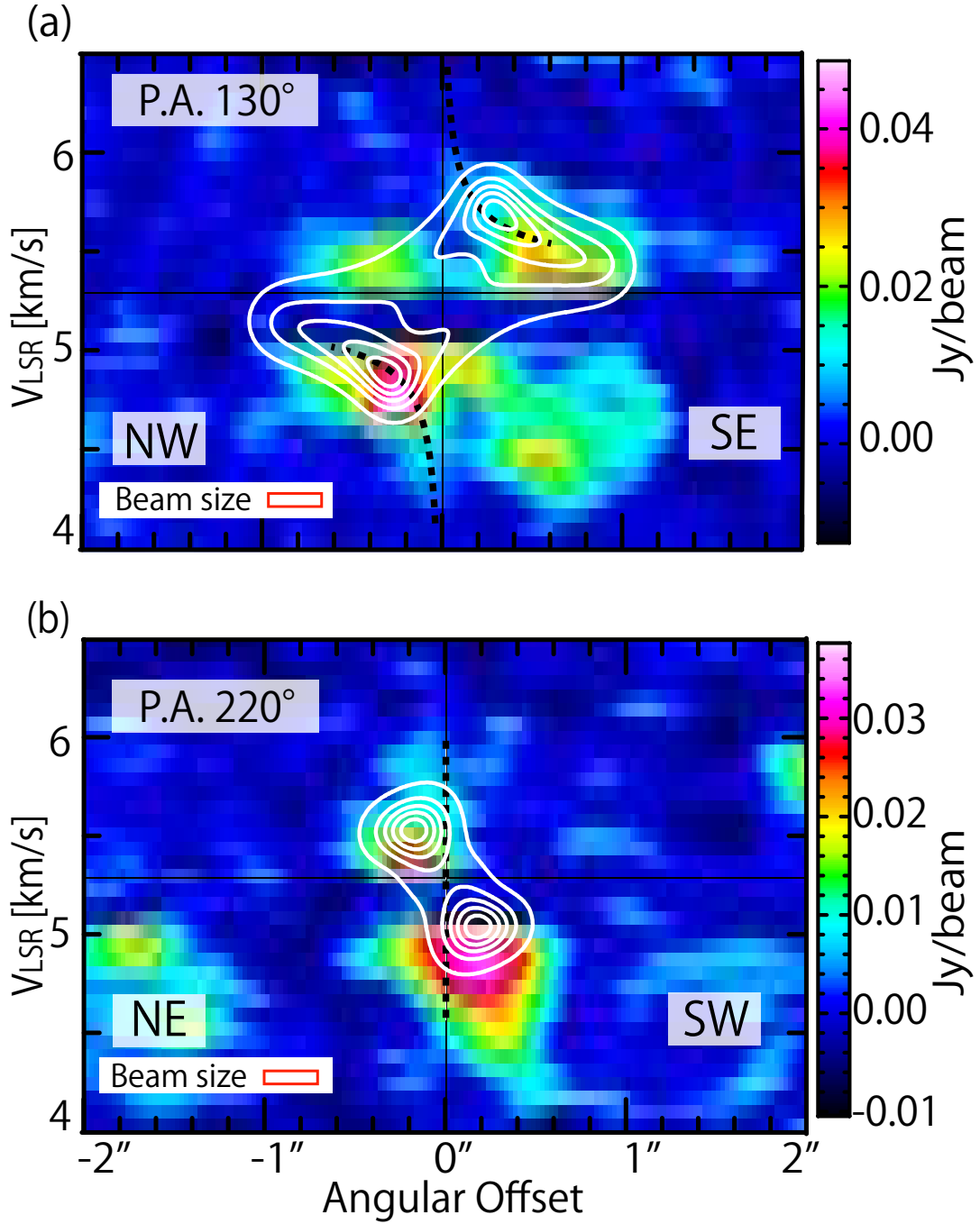}
\caption{The PV diagram of the CCH ($N=3-2, J=7/2-5/2, F=4-3$) emission along the envelope direction (a), and along the line perpendicular to the envelope direction (b). The dashed lines show the model of Keplerian rotation with the protostellar mass of 0.007 $M_\odot$. The white contours are the results of the infalling-rotating envelope model with a protostellar mass of 0.007 $M_\odot$, the outer radius of 155 au, the radius of the centrifugal barrier of 40 au, and inclination angle of 70\degr. \label{Figure3}}
\end{figure}

\begin{figure}[h]
\centering
\includegraphics[scale=0.75]{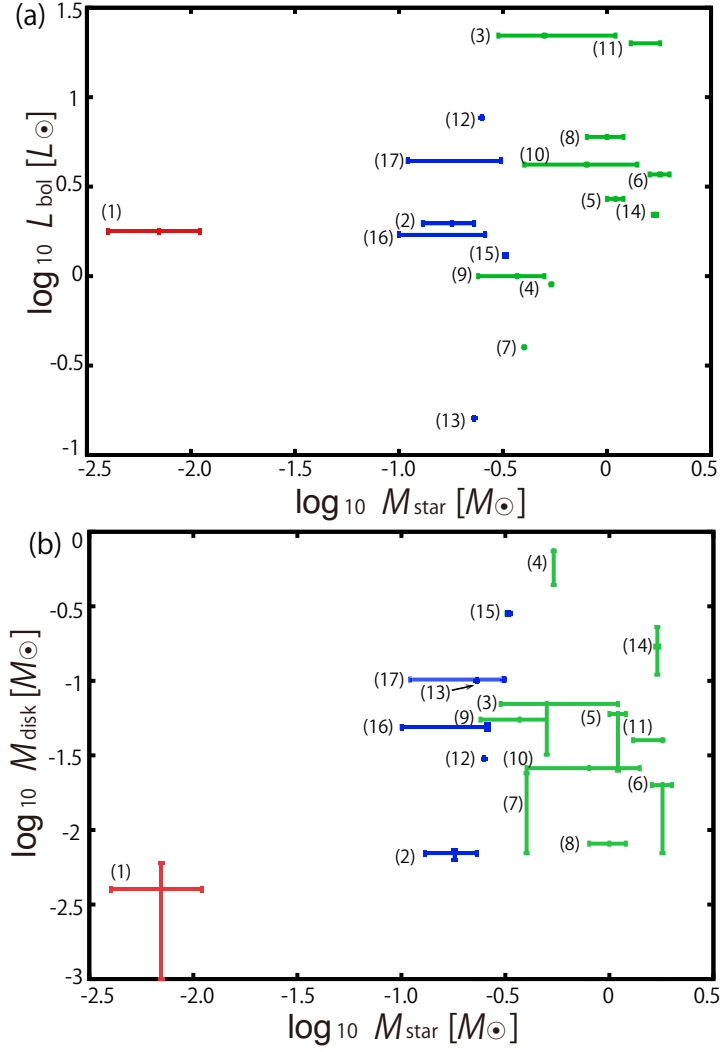}
\caption{(a) Comparison between the protostellar masses and the bolometric luminosities. (b) Comparison between the protostellar masses and the disk masses. They are for the sample of protostars in previous studies listed below. Red marks with error bars represent IRAS 15398-3359 (this study). Blue and Green marks show the protostars of $T_{\rm bol}< 70$ K (Class 0) and $T_{\rm bol}> 70$ K (Class I), respectively. Error bars show the ranges of the mass. The sources used in the plots are as follows: (1)IRAS 15398-3359 \cite[this work]{Jorgensen et al.(2013)} (2) L1527 IRS \citep{Kristensen et al.(2012), Tobin et al.(2012), Sakai et al.(2014a)} (3) L1551 IRS5 (4) TMC1 (5) TMC1A (6) L1489 IRS (7) L1536 (8) IRS 43 (9) IRS 63 (10) L1551NE \citep{Chou et al.(2014)} (11) HH111 \citep{Lee(2011),Lee et al.(2018)} (12) HH212 \citep{Lee et al.(2017)} (13) Lupas3 MMS \citep{Yen et al.(2017)} (14) BHB07-11 \citep{Alves et al.(2017), Yen et al.(2015a)} (15) IRAS 03292+3039  (16) L1448 IRS2 (17) L1448C \citep{Tobin et al.(2015), Yen et al.(2015a)}. In some protostars, only the range of the protostellar mass and that of the disk mass are estimated. In this case, we employ the maximum range. \label{Figure4}}
\end{figure}


\end{document}